\documentclass[12pt]{iopart}
\usepackage{iopams}  

\usepackage{graphicx}

\usepackage{bm}

\renewcommand{\tabcolsep}{4pt}

\begin{document}
	
	\title{Multi-layer Trajectory Clustering: A Network Algorithm for Disease Subtyping}
	
	\author{Sanjukta Krishnagopal}
	\address{Departments of Physics, University of Maryland, College Park, Maryland, USA 20742.}
	\address{Gatsby Computational Neuroscience Unit, University College London, London, United Kingdom W1T4JG.} 
	\ead{sanjukta@umd.edu }
	\vspace{10pt}
\date{\today}
\begin{abstract}
Many diseases display heterogeneity in clinical features and their progression, indicative of the existence of disease subtypes. Extracting patterns of disease variable progression for subtypes has tremendous application in medicine, for example, in early prognosis and  personalized medical therapy. This work presents a novel, data-driven, network-based Trajectory Clustering (TC) algorithm for identifying Parkinson's subtypes based on disease trajectory. Modeling patient-variable interactions as a bipartite network, TC first extracts communities of co-expressing disease variables at different stages of progression. Then, it identifies Parkinson's subtypes by clustering similar patient trajectories that are characterized by severity of disease variables through a multi-layer network. Determination of trajectory similarity accounts for direct overlaps between trajectories as well as second-order similarities, i.e., common overlap with a third set of trajectories.
This work clusters trajectories across two types of layers: (a) temporal, and (b) ranges of independent outcome variable (representative of disease severity), both of which yield four distinct subtypes. The former subtypes exhibit differences in progression of disease domains (Cognitive, Mental Health etc.), whereas the latter subtypes exhibit different degrees of progression, i.e., some remain mild, whereas others show significant deterioration after 5 years.
The TC approach is validated through statistical analyses and consistency of the identified subtypes with medical literature. This generalizable and robust method can easily be extended to other progressive multi-variate disease datasets, and can effectively assist in targeted subtype-specific treatment in the field of personalized medicine.

\end{abstract}


\maketitle

\section{Introduction}

The field of network medicine has gained tremendous traction in recent years. With the growth of public disease datasets and development in computational methods, several recent works have attempted to use data-driven methods to study disease progression \cite{OXT18,KHO19}. Network medicine \cite{6,8,9,10,SON19} offers innovative data-driven solutions to disease modeling through analyzing complex interactions within data. For instance, \cite{10} identifies disease-related genetic variants through network-based studies of the human disease network (i.e. the ‘diseaseome’), in which diseases are linked if they share one or more associated genes. Functional brain networks for neurobiologically relevant graphical properties of Alzheimers disease are analyzed in \cite{JAL17}. However, network-based patient subtyping based not only on disease variable values but also on their trajectories, i.e., evolution of variable patterns, is relatively unexplored. 
 
Parkinson’s Disease, a degenerative neurological disorder, is the second most common neurodegenerative disorder following Alzheimer’s disease, affecting an estimated 7-10 million people worldwide \cite{1}. It is both chronic, meaning it persists over a long duration, and progressive, meaning symptoms, such as tremor, loss of memory, impaired balance etc. worsen with time \cite{SCH06,LEW05}. In addition, it is a highly variable disease, with rate and type of progression differing significantly across the population \cite{2}. It is increasingly evident that Parkinson's disease is not a single entity but rather a heterogeneous disorder with multiple subtypes. Several studies have attempted to classify patients into subtypes \cite{KRI20, SAU16, VAN11, SEL09}.Two recent studies have developed models of PD progression based on clinical, demographic and genetic data at baseline, using hierarchical cluster analysis \cite{5} and a Bayesian multivariate predictive inference platform \cite{16}, to identify PD subtypes. However, there is no overall consensus on Parkinson's subtypes \cite{5}, and little is known about the effect of the interplay between different types of variables on Parkinson's progression. Early recognition of patient subtype allows medical workers to employ subtype-specific treatment, potentially improving and prolonging life \cite{MAR13,THE14}. However, current approaches in Parkinson's subtyping have largely not explored data-driven analyses to unravel multiple complex influences of a large number of longitudinal variables on disease progression \cite{24}.

Sophisticated data-driven methods are required to extract meaningful information effectively from medical datasets. In recent years, several deep learning approaches to disease-subtyping have emerged \cite{ZHA19,MIO18}. However, such approaches are limited in extracting interpretable information about the micro-structure of the subtypes, i.e., the variable relationships that underlie the subtype. In contrast, network science approaches offer an intuitive visualization for modeling relationships between different types of variables. Additionally, evolving interactions between variables are easily represented through a multi-layer structure.

Significant research has been done in clustering in a multi-layer network \cite{DONG12,KIM15}. However, subtyping through trajectory clustering is relatively unexplored in network medicine\cite{14,15}. This work presents a novel multi-layer-network-based Trajectory Clustering (TC) algorithm to identify disease subtypes based on similarities in trajectories through variable clusters. First, it identifies variable-communities (clusters of variables that co-express) through modeling patient-variable interactions as a bipartite network. It then tracks patient-memberships through multiple layers of variable-communities to define their trajectories. Lastly, disease subtypes are identified by clustering similar trajectories. To the best of the author's knowledge, the only other work with a trajectory-based approach to Parkinson's subtyping is the author's previous work \cite{KRI20}. While their research questions and data are the same as those in this paper, \cite{KRI20} defines trajectories as matrices (and implements a trajectory \textit{profile} clustering) which is fundamentally different than the method proposed in this paper, and yields different, although complementary, results. Additionally, it is a first-order method. In contrast, this paper defines trajectories through a multi-layer network and identifies subtypes using a graph-based second-order trajectory clustering. The novel contributions of this approach are multifold: (1) a unique subtyping approach on graphs with intuitive visualization of trajectories, (2) identification of co-expressing variable clusters in the stacked-multi-layer network offering a variable-centric perspective to disease progression (3) studying disease progression as a function of \textit{any} outcome variable (with results presented for outcome variable MDS-UPDRS3). The contributions of this work are distinct from and complementary to those in \cite{KRI20}.

The contributions of this work are (a)  extracting clusters of co-expressing variables at different stages of disease, and (b) identifying subtypes based on similarities in disease trajectories. The set of variable clusters identified by our network are in agreement with Parkinson's disease domains recognized in medical literature \cite{MAR18}. Disease subtypes identified are shown to be statistically distinct and are supported by the results in \cite{KRI20}. This easily generalizable approach for multi-layer trajectory clustering presents a unique way for extracting patterns of disease progression in complex longitudinal medical data, and helps bridge the gap between data-based computational approaches and applied medicine.

\section{Data}

\begin{figure}[htbp!]
     \begin{center}
             \includegraphics[width=0.95\linewidth]{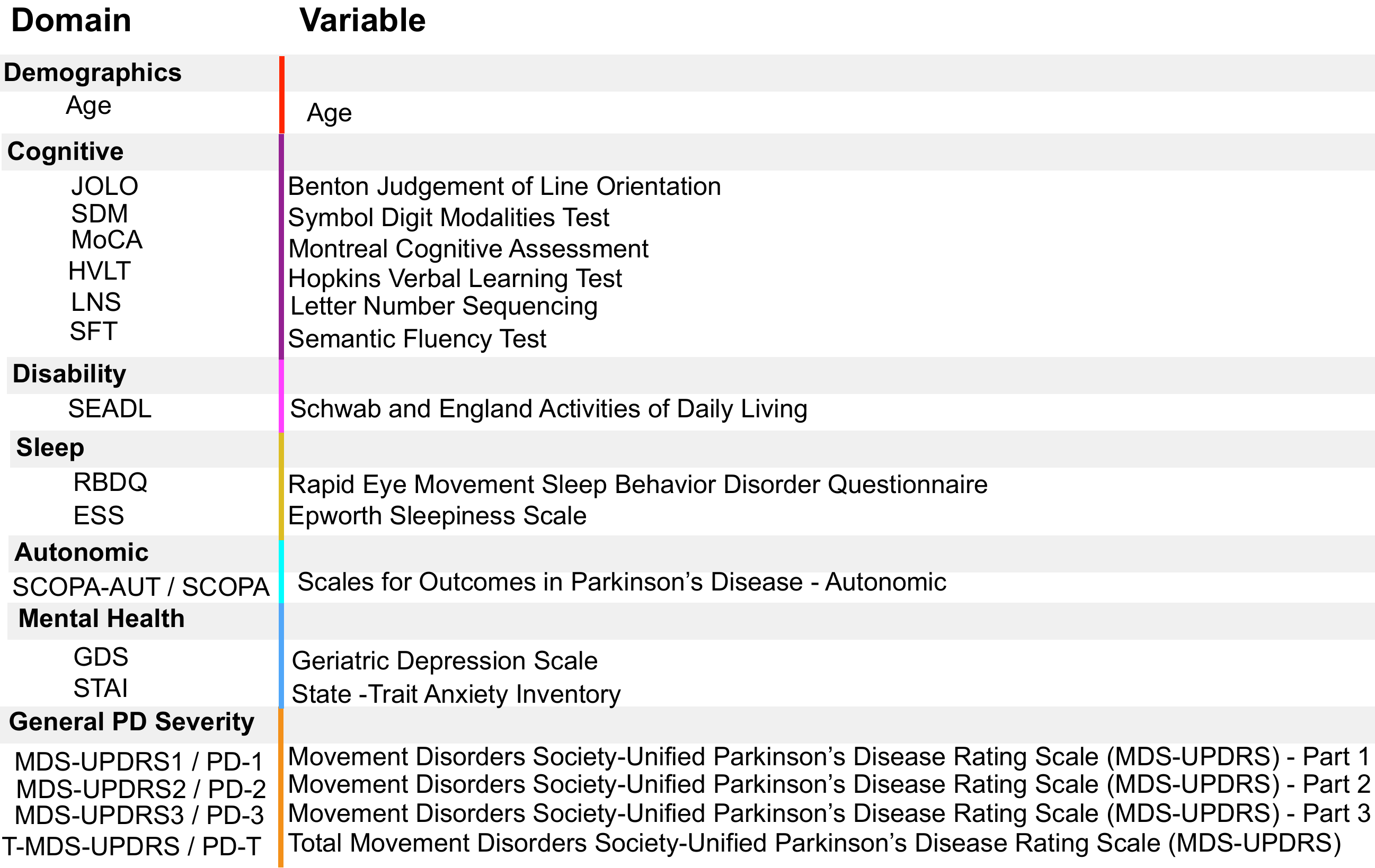}
    \end{center}
    \caption{Description of the six medical domains and their variables in the PPMI clinical dataset.}
    \label{data}
\end{figure}

Data used in the preparation of this article was obtained from the Parkinson’s Progression Markers Initiative (PPMI) database (www.ppmi-info.org/data). It consists of patient variable values across 5 timepoints: baseline year (denoted as year0) and years-1,2,3, and 4. Upon excluding patients with incomplete data, 194 patients remained in the analysis.The dataset categorizes the clinical variables into domains (Cognitive, Behavioral, Sleep, PD Severity, Autonomic, Disability) based upon functionality as shown in \ref{data}. Variable data is obtained from commonly used standard medical tests, a few of which are outlined below. JOLO is a standardized test of visuospatial skills. SDM is a test measuring neurological information processing speed and efficiency using a symbol/digit substitution tasks. HVLT is a word-learning test that measures episodic verbal memory and recall ability. SEADL assesses the difficulties patients have completing daily activities or chores. EPS test scores give a subjective measure of a patient's daytime sleepiness. STAI is used in clinical settings to diagnose anxiety and to distinguish it from depressive syndromes. The MDS - UPDRS scale includes series of ratings for typical Parkinson’s symptoms that cover all of the movement hindrances of Parkinson’s disease and consist of Mentation, Behavior and Mood, Activities of Daily Living, Motor sections etc.

According to PPMI, motor assessment for variables in this dataset was performed in a `practically defined off' state, i.e., subjects are asked to withhold medication prior to the assessment for 12 hours, practically eliminating medication effects on motor symptoms.

\section{Network Architecture: Stacked Multi-layer Network}
\label{bip}

Variable-communities within each layer are identified and stacked to create a multi-layer framework across which patient trajectories are defined. This proceeds as follows: 

\begin{enumerate}
\item Within each layer, individual-variable relationships for $I$ patients and $V$ relationships, are modeled as a bipartite network. Bipartite networks are graphs consisting of two sets of disjoint and independent nodes, such that there exist no edges between nodes belonging to the same set. They are a natural choice for modeling relationships between two different classes (in this case patients variables). The adjacency matrix $Z$ of the bipartite network is created  as follows:
\begin{itemize}
\item The `direction' of each variable is determined. If disease progression (how severely a patient is affected) is positively correlated with values of the disease variable $v$) then the direction of the variable $d_v$ is set as +1, and if disease-affection is negatively correlated with disease variable values, then $d_v = -1$. These directions are standard and well known for variables in the PPMI clinical data. Variables ESS, RBDQ, GDS, STAI, MDS-UPDRS-1,2,3,T and Age have $d_v =1$ and HVLT, JOLO, SFT, LNS, SDM, MoCA, SEADL have $d_v =-1 $. 
\item  The value ($F_{iv}$) of variable $v$ in individual $i$ is then converted to a z value, $z_{iv}$ that normalizes it between $0$ and $1$,
\begin{equation}
z_{iv} =\frac{F_{iv} - \min_{j \in I}F_{jv}}{\max_{j \in I}F_{jv}-\min_{j \in I}F_{jv}}.
\end{equation}	
These z values are continuous and non-thresholded.
\item The adjacency matrix $Z$ of size $I \times V$ for the patient-variable bipartite network is populated as
\begin{equation}
Z_{ij} =
\cases{
 z_{iv} 	 \quad  &if  $d_v=1$ \\
1-z_{iv}   \quad  &if $d_v=-1$.}
\end{equation}
\end{itemize}
\item An independent patient-variable bipartite network is generated for each layer (an illustrative bipartite network is shown in figure \ref{mult_bip} (left)), where a layer could represent a time-point or a range of outcome variable values. An outcome variable is simply any variable as a function of which one may desire to measure disease progression as indicated by non-outcome variables. This paper presents results for outcome variable chosen to be MDS-UPDR3. Then Louvain community detection \cite{18} is implemented on the bipartite network for each layer $l$. Louvain community detection yields the number of communities through optimization of the Newman-Girvan modularity function \cite{19} and has no hyperparameters. It identifies variable-communities $C_k^l$, where $k \in [0, \ldots, K_l-1]$, comprising of patients and variables (shown through shaded ovals in figure \ref{mult_bip} (left)), and $K_l$ is the total number of variable-communities in layer $l$. 

\begin{figure}[htbp!]
	\begin{center}
		\includegraphics[width=0.95\linewidth]{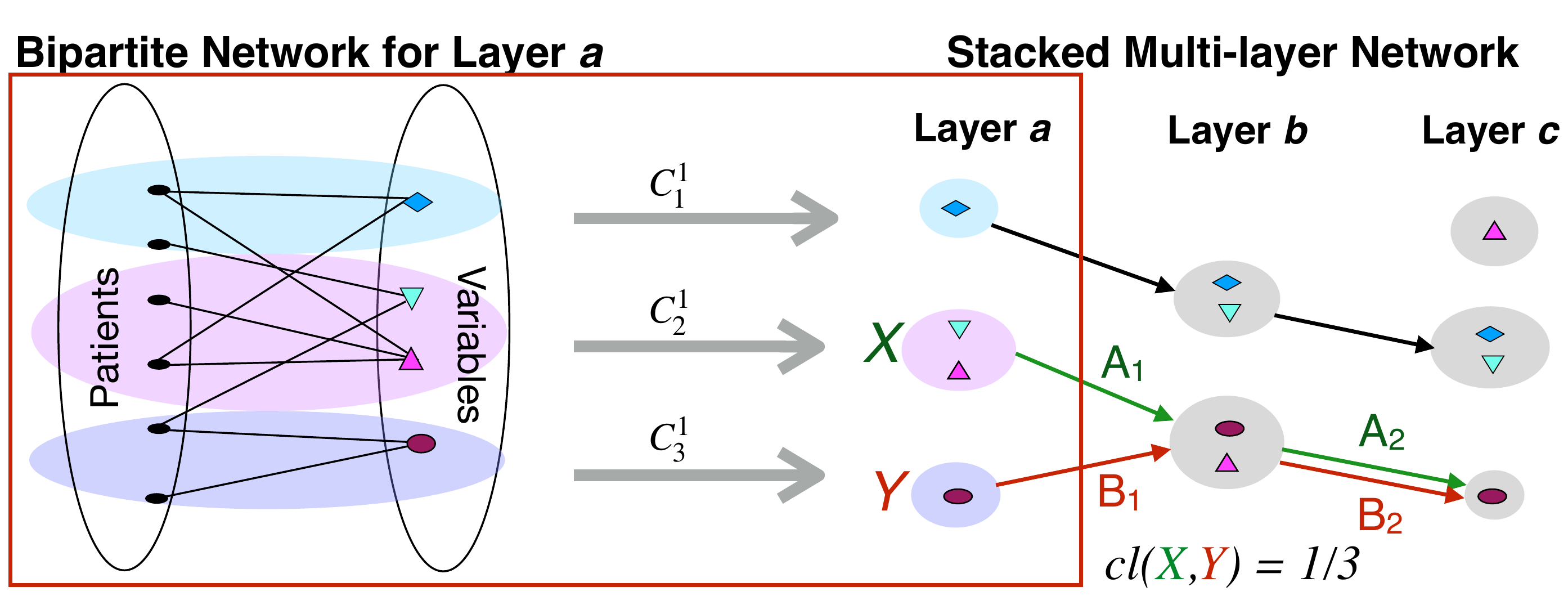}
	\end{center}
	\caption{(left) shows a representative bipartite patient-variable network for one layer. The highlighted ovals represent variable-communitites consisting of patients and disease variables. (right) represents a stacked multi-layer graph over three layers, where the variable-communities from (left) form the first layer. Three sample trajectories are shown, and the corresponding closeness between nodes $X$ and $Y$ is calculated.}
	\label{mult_bip}
\end{figure}

The TC algorithm to cluster patient trajectories (and hence patients) is presented in sec. \ref{tc_bip}.

\item These variable-communities in each layer, variable-community represented as a node, are stacked across all layers as shown in figure~\ref{mult_bip}(right). This is the stacked multi-layer graph ($G$). Each individual belongs to a variable-community $C_k^l$ at each time-point. One can then track individual $i$ membership over time to define a patient trajectory $T_i$.
\end{enumerate}

\section{Trajectory Clustering Algorithm}

\label{tc_bip}
Each node in $G$ is a variable-community. Trajectory Clustering (TC) is a second-order algorithm identifies patient subtypes based by clustering patients with similar trajectories of disease progression on the stacked multi-layer network $G$ as follows:

\begin{enumerate}
\item `Node closeness' $cl(X,Y)$ between two nodes $X$ and $Y$ in graph $G$ is defined as the fraction of whole or partial trajectories passing through $X$ and $Y$ that overlap. Suppose $A$ and $B$ are sets of all edges (both inter-layer and intra-layer), with repetition, that belong to trajectories passing through node $X$ and node $Y$ respectively, then node closeness between is defined as
\begin{equation}
cl(X,Y)=\frac{\#(A \cap B )}{\#( A \cup B)}.
\end{equation}
where $\#$ denotes the size of a set. Note that $cl(X,X)=1.$ Figure~\ref{mult_bip} (b) presents a simple example of node closeness calculated between nodes $X$ and $Y$. There is one overlapping edge-pair {$A_2, B_2$} on trajectories passing through nodes $X,Y$, and the total  size of the set of edges on these trajectories is $3$. Hence, $cl(X,Y)=1/3$.

\item Trajectory Similarity ($\eta(T_i,T_j)$) between two individual (patient) trajectories $T_i$ and $T_j$ is then given by $\Sigma_{t}cl(X(t)_{T_i},Y(t)_{T_j})$,  where $X(t)_{T_i}$ is the node $X(t)$ that trajectory $T_i$ passes through at timepoint $t$. $\eta(T_i,T_j)$ measures not just first-order similarities between two trajectories, but also second-order similarities i.e., interactions between two non-overlapping trajectories that result from their mutual overlap with a third set of trajectories. Important properties of the second-order TC algorithm are outlined in \ref{timex}.

\item A patient-patient network $P$ is created whose adjacency matrix is given by
\begin{equation}
P_{i,j}= \eta(T_i,T_j).
\end{equation} .
The network P contains information of how similar patients are in terms of their disease trajectories. $P_{i,j}$ gives the trajectory similarity $\eta(T_i,T_j)$ between patients $i$ and $j$. Thus a higher value of $P_{i,j}$ indicates that patients $i$ and $j$ are connected with a higher weight in the network $P$ and are correspondingly more likely to have similar disease progression patterns. $P$ is a symmetric matrix with $P_{i,i}=1$. 
\item Louvain community detection is implemented on the network P (for 100 runs), and the maximally occuring communtiy configuration over all runs was chosen to obtain $M$ trajectory-communities $S_k$ (where $k \in [0, \ldots M-1]$) with similar disease progression trajectories. In a modularity maximizing algorithm such as Louvain \cite{18}, the optimal number of trajectory-communities are unconstrained and automatically identified to correspond to the configuration with maximal modularity (a commonly used metric for community detection in networks).
\end{enumerate}



\section{Results}

\subsection{Bipartite variable-communities}

\begin{figure*}[ht!]
	\begin{center}
		\includegraphics[width=0.95\linewidth]{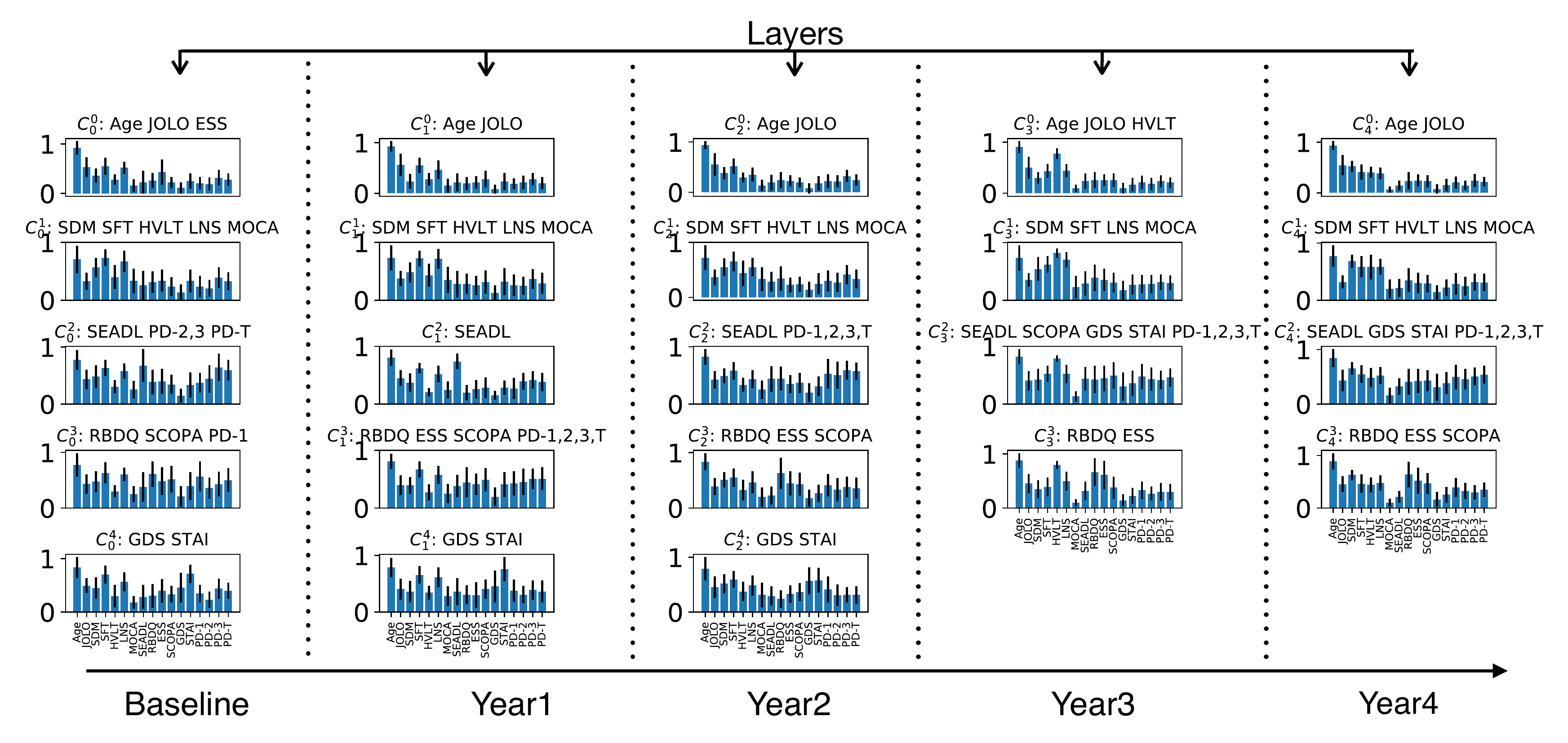}
	\end{center}
	\caption{Community profiles of the variable-communities in each layer. Here, the layers represent time-points. Each panel represents one variable-community comprising of individuals and variables. These communities are characterized by the variables in them, listed above each panel. The panels show the average $z$ value of all people in that variable-community, and error bars are given by their standard deviation.}
	\label{multi}
\end{figure*}

Upon creating an individual-variable network for each layer (where each layer is a timepoint) as in figure \ref{bip} (left), Louvain community detection is performed on each layer. Figure ~\ref{multi} shows the communities $C_l$ identified in each layer $l$ comprising of individuals as well as dominant variable. These variable-communities are characterized by the dominant disease variables that belong to them (listed above each variable-community in figure~\ref{multi}). 

The variable-communities are indicative of sets of variables that patients co-express with high severity in different stages of the disease. For instance, the last panel (bottom row) in the baseline year indicates that in early stages of disease, patients displaying severity in variable GDS for instance, are also likely to display severity in STAI. 
While variables are treated independently in this analysis, the emergent variable-community structure is largely consistent with domain structure in medical literature \cite{MAR18}. The composition of variable-communities remains similar across the layers. Age and JOLO (a cognitive variable) cluster together in all layers. Another variable-community comprises of all the other cognitive variables. Disability (SEADL) and General PD Severity Variables (PD-1,2,3,T) commonly co-express with high severity together in all layers except in the baseline year where PD-1 is found to co-express with Autonomic (SCOPA-AUT) and Sleep (RBDQ) variables. As  General PD Severity variables represent motor functioning, their co-occurence with disability is expected.  Sleep and Autonomic variables often clustered together in several layers. Mental Health variables (GDS, STAI) initially form an independent variable-community, however, as time progresses, they co-express with high severity with PD Severity and Disability domains. Deterioration of mental health is expected with worsening motor skills and increased disability, particularly in later stages of disease progression. The clustering of variables in the same domain is further validation of our method yielding medically relevant results. The individuals within each variable-community have relatively high average severity of corresponding variables that characterize the community. 

\subsubsection{Statistical analysis}

The Kruskal-Wallis statistical test \cite{KRU52} is an estimate of whether two variables are sampled from the same distribution. Kruskal-Wallis test for multiple groups was conducted to validate this approach and demonstrate some of the differences between variable-communities. This test is chosen because it is applicable in cases such as this where values for several of the variables violate the normality assumption. 

\begin{table*}[htbp!]
		\caption{\label{stat1} Comparing the variable-communities (described in figure~\ref{multi}) in each layers (representative of timepoints) through the Kruskal-Wallis p-values. }
\begin{indented}
	\item[]
\lineup
	\begin{tabular}{ ccccccc } 
	\br
		Variable & Median & Year0 &  Year1 & Year2 & Year3 & Year4  \\ 
	\mr  
		
		Age & 71.00&1.759E-08 & 2.439E-07 & \textbf{2.263E-09} & 2.966E-06 & 5.115E-08 \\
		 \textbf{Cognitive} &&&&&&\\
		JOLO* & 14.00& 3.687E-06 & 6.807E-06 & 8.491E-05 & 2.769E-03 & 1.526E-09 \\
		SDM* & 42.00&3.951E-07 & 3.764E-09 & 7.653E-07 & 5.050E-08 & 7.435E-12 \\
		SFT* & 47.50&5.151E-07 & 1.920E-06 & 1.381E-04 & 3.276E-09 & 2.753E-07 \\
		HVLT* & 0.90& \textbf{3.222E-03} & 6.110E-07 & \textbf{5.731E-03} & \textbf{2.959E-01} & 7.282E-05 \\
		LNS* & 11.00&1.207E-05 & 1.457E-08 & 1.347E-07 & 1.234E-11 & 5.982E-11 \\
		MOCA* & 27.00& 1.767E-05 & 3.470E-05 & 1.163E-06 & 2.379E-04 & 2.928E-08 \\
		\textbf{Other} &&&&&&\\
		 SEADL* &95.00& 4.317E-10 & 5.634E-10 & 1.039E-07 & 6.277E-07 & 3.821E-08 \\
		RBDQ & 4.00&2.931E-08 & 1.247E-06 & 1.212E-09 & 2.699E-10 & 1.172E-11 \\
		ESS & 6.00& \textbf{1.265E-01} & 6.166E-08 & 6.262E-08 & 4.623E-10 & 1.191E-08 \\
		SCOPA-AUT & 9.00& 6.022E-09 & 4.087E-07 & 5.470E-10 & 8.695E-09 & 2.638E-10 \\
		GDS & 2.00&6.678E-06 & 1.323E-08 & 7.281E-05 & 2.941E-08 & 9.002E-09 \\
		STAI & 65.00&1.805E-07 & 4.617E-09 & 1.374E-03 & 6.388E-07 & 2.223E-08 \\
		\textbf{General PD} &&&&&&\\
		PD-1 & 5.50&3.609E-10 & 8.432E-09 & 1.789E-09 & 3.372E-12 & 4.205E-12 \\
		PD-2 &5.00& 1.471E-08 & 1.969E-09 & 1.231E-08 & 1.599E-12 & 2.570E-15 \\
		PD-3 & 20.00&2.587E-08 & 1.088E-08 & 1.026E-10 & 2.275E-09 & 4.794E-12 \\
		PD-T &32.00& 1.674E-14 & 5.236E-14 & 1.280E-14 & 9.158E-16 & 5.981E-17\\
			\br
\end{tabular}
\end{indented}

\end{table*}

Table~\ref{stat1} tabulates the medians and the p-values of the Kruskal-Wallis statistical test applied to each layer. Medians are calculated from the whole population raw data. Variables with negative directions are denoted by an asterisk (*). 
The p-values are a measure of statistical differences amongst the variable values of the individuals belonging to the different variable-communities. To account for Type I errors due to multiple comparisons the Benjamini-Hochberg False Discovery Rate method \cite{BEN95} is used. This gives us an adjusted significance level for each of the p-values $\alpha_{adjusted} = \frac{\alpha \times i}{n_c}$, where $n_c$ is the total number of comparisons, and $i$ is the rank of the p-value (for example, the smallest has a rank of 1, the second smallest has a rank of 2 etc.). Significance value $\alpha = 0.05$. Total number of comparisons  $n_c = T \times V = 5 \times 17 = 85$, where $T$ is the number of timepoints, and $V$ is the number of variables. For instance, in Table~\ref{stat1}, HVLT in year 3 has the largest p-value (rank one), hence it has the strictest acceptance threshold given by $\alpha_{adjusted} = \frac{0.05 \times 1}{85} = 5.882 \times 10^{-3}$. Comparisons not meeting the criteria for statistical significance are shown in bold text. A majority of the p-values are below their adjusted significance level, suggesting that there is an underlying statistical difference between the variable distributions in the variable-communitites in a layer.

\subsection{Trajectory Clustering on Temporally Stacked Multi-layer Network}
\label{timex-cluster}

\subsubsection{Trajectory communities}

Each individual is a member of one of the variable-communities at each time-point (or in each layer in this case). Tracing their membership yields their trajectory. The TC algorithm (sec. \ref{bip}) identifies patient subtypes based on similarities in their trajectories through the multi-layer network. The run time of the trajectory clustering algorithm on the temporally stacked multi-layer network is $4.9346s$. All code was run on a laptop with 2.8 GHz Intel Core i5 processor and 8 GB RAM.

\begin{figure*}[htbp!]
     \begin{center}
             \includegraphics[width=0.95\linewidth]{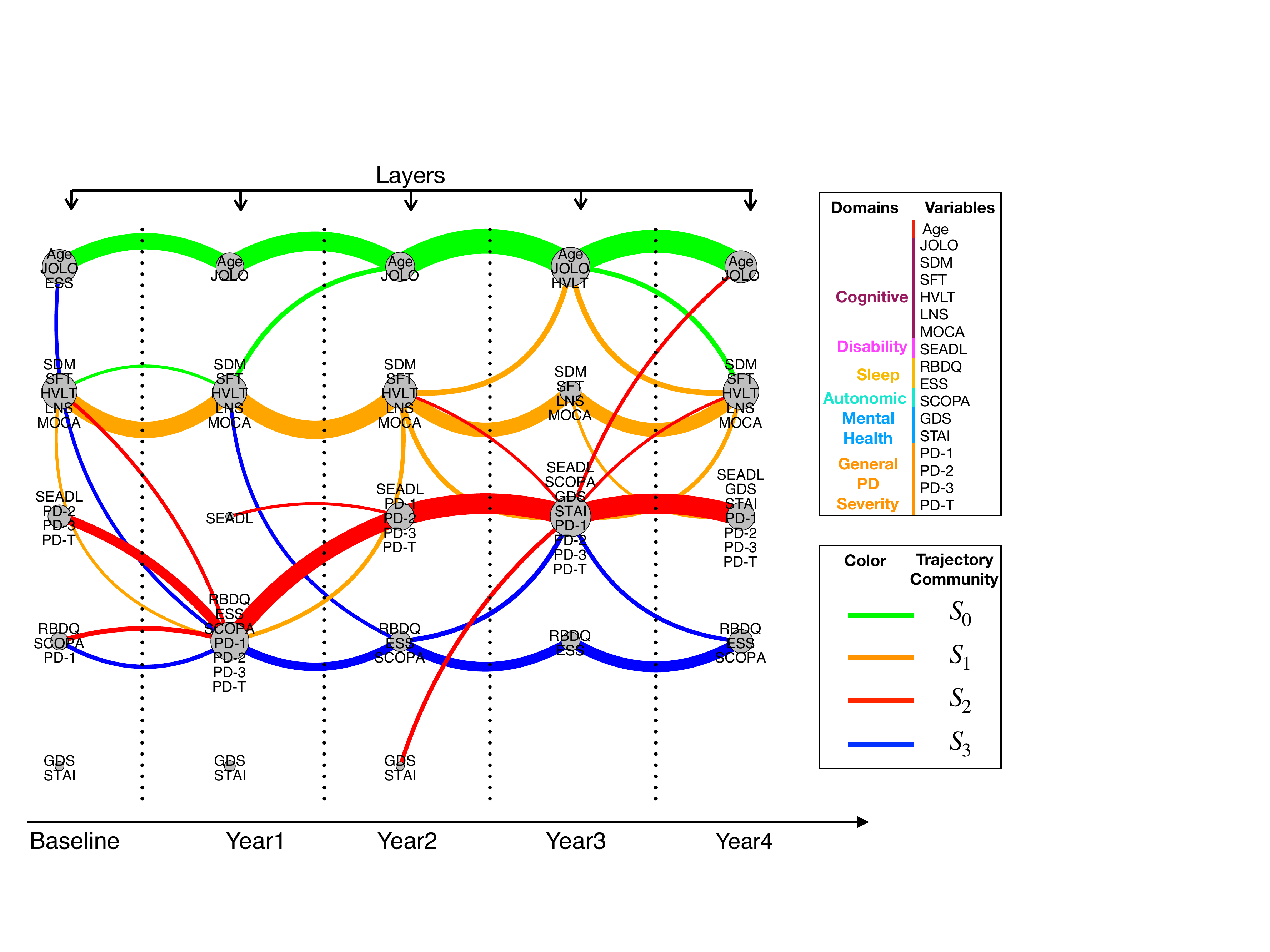}
    \end{center}
    \caption{Each node represents a variable-community consisting of variables and patients. The size of the node denotes the number of people in the variable-community. The variables are color-coded by domain. Patient trajectories from the baseline layer to the year4 layer are clustered using the TC algorithm. The trajectory-communities (subtypes) are color-coded and directed (from left to right). The thickness of a colored edge denotes the number of patients flowing along that edge in the corresponding trajectory-community. Number of people in each subtype are as follows: green - 56, orange - 55, red - 53, blue - 33. Edges with fewer than 3 people are not plotted.}
    \label{TS_bip}
\end{figure*}

As shown in figure~\ref{TS_bip}, the TC algorithm identifies four distinct trajectory-communities (subtypes) for disease progression across temporal layers. The edges are directed from left to right (baseline year to year4), shown without arrows for ease of viewing. The thickness of the edges is a measure of the flow of people between the corresponding variable-community nodes. Thus, edge thickness is an estimate of the probability of transition of patients expressing high severity from one variable set to another as time progresses.

The green subtype $S_0$ with 56 patients is characterized by an older population and higher severity of JOLO, as well as increase in general Cognitive impairment as time progresses. The orange subtype $S_1$ with 55 patients displays high values of Cognitive variables. In addition, it progresses to high severity in General PD Severity and Sleep variables. The red subtype $S_2$ with 53 patients is characterized by high Disability and severe General PD (motor), and develops severity in Mental Health variables through progression of time. Lastly, the blue subtype $S_3$, the smallest trajectory-community with 33 patients, is characterized by severe Autonomic and Sleep variables. It shows limited Cognitive and General PD severity in earlier years, and the severity of expression of these variables reduces over time.

\subsubsection{Trajectory-community profiles}

\begin{figure*}[ht!]
	\begin{center}
		\includegraphics[width=0.95\linewidth]{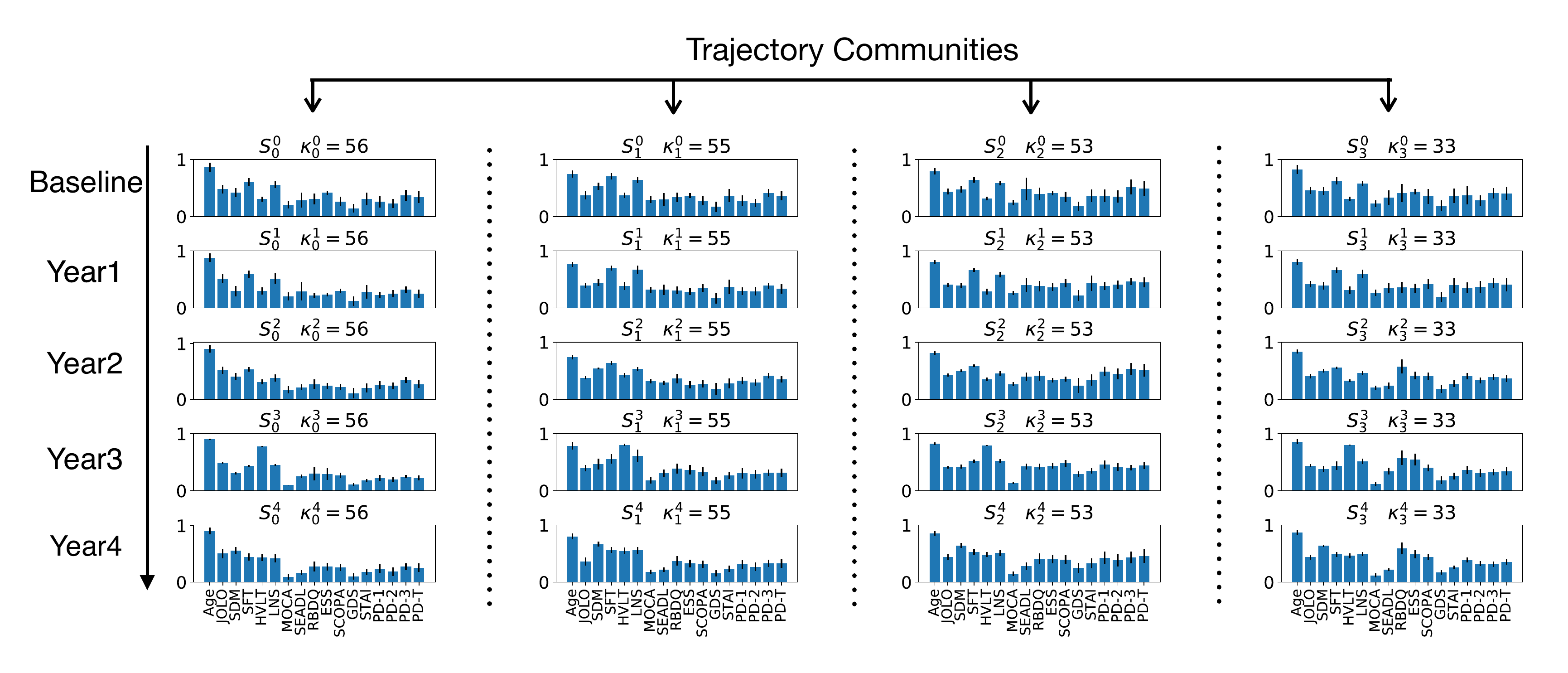}
	\end{center}
	\caption{Trajectory profiles of the 4 trajectory-communities. The $i^{th}$ vertical set of panels corresponds to a trajectory-community denoted by $S_i$). Within each column, the five panels arranged from top to bottom show the average profile in years 0(baseline),1,2,3,4 respectively. Each individual panel shows the average $z$ values across all variables of $\kappa_i^l$ patients in a trajectory-community.}
	\label{v_prof}
\end{figure*}

Figure ~\ref{v_prof} shows the trajectory profile (weighted average profile of all individuals belonging to a trajectory-community) across all layers (which is equivalent to timepoints in this case). The trajectory profiles ($S_i^t$) of  trajectory-community $i$ at timepoint $t$ is obtained by taking the weighted average of the variable-communities that members of the trajectory-community belong to as follows:
\begin{equation}
S_i^t=\frac{\Sigma_{i} X(t)_{T_i}}{\kappa_i^t}.
\label{tc_pr}
\end{equation}
where $i \in \kappa_i^t$ and $\kappa_i^t$ is the total number of individuals in trajectory-community $S_i$ and layer $t$. $X(t)_{T_i}$ is the variable-community that individual $i$ belongs to at time $t$. As seen in figure~\ref{v_prof}, each trajectory-community has a distinct trajectory profile with different evolution patterns and high values of their corresponding dominant variables.

The green trajectory-community $S_0$ with 56 members has a relatively mild disease stage, despite having a relatively older population. Disability, Autonomic, and General PD severity start very low, and continue to stay low over time. Amongst the cognitive variables, JOLO is slightly higher than in other trajectory-communities through the layers, ESS starts high and reduces over time, whereas HVLT gets more severe. 
The orange trajectory-community $S_1$ with 55 members show consistently high severity in all Cognitive variables except JOLO and MoCA. Other variables are low; Disability reduces across the layers, whereas Sleep variables show a slight increase as disease progresses across time.
 The red trajectory-community $S_2$ with 53 members has relatively severe General PD Severity motor variables. As time progresses, Autonomic, Mental Health and Disability demonstrate growth in severity. 
 Lastly, the blue trajectory-community $S_3$, the smallest with 33 members show heterogeneity in the variable structure. General PD Severity variables remain consistently low through time, whereas Sleep variables RBDQ and ESS show growth in severity across the layers.
 
\subsubsection{Statistical analysis}

To validate the algorithm and demonstrate the differences in the \textit{progression} of the different trajectory-communities, the Kruskal-Wallis statistical test for multiple groups was conducted on the \textit{difference} between the consecutive year variable values. This test is chosen because it is applicable in cases such as this where values for several of the variables violate the normality assumption. We do not include the variable `age' in the statistical analysis since the difference in age between consecutive years is exactly 1 regardless of the subtype.

\begin{table}
		\caption{ \label{stat2}
		Comparing evolution of trajectory-communities by calculating the Kruskal-Wallis p-values for variable differences between consecutive years.}
	\begin{indented}
		\item[]
		\lineup
	\begin{tabular}{ cccc } 
		\br
	Variable & Year1-Year0 & Year2-Year1 & Year3-Year2  \\ 
	\mr
	\textbf{Cognitive} &&&\\
	 JOLO* & 1.451E-06 & 7.295E-04 & 3.335E-11 \\
	 SDM* & 2.579E-05 & \textbf{2.556E-02} & 6.247E-05 \\
	 SFT* & 2.089E-05 & 9.683E-07 & 1.005E-05 \\
	 HVLT* & 4.225E-04 & 1.892E-05 & 1.548E-18 \\
	 LNS* & 3.724E-06 & \textbf{8.711E-02} & 3.486E-04 \\
	 MOCA* & 2.562E-04 & 5.158E-07 & 3.142E-13 \\
	 \textbf{Other} &&&\\
	 SEADL* & \textbf{1.092E-02} & 1.809E-07 & 5.618E-08 \\
	 RBDQ & 9.470E-04 & 3.796E-09 & \textbf{2.015E-02} \\
	 ESS & 6.881E-15 & 1.059E-07 & 1.708E-04 \\
	 SCOPA-AUT & 3.035E-03 & 1.202E-05 & 3.529E-11 \\
	 GDS& 4.778E-08 & 6.070E-03 & 4.111E-05 \\
	 STAI & 3.209E-07 & \textbf{4.996E-02} & 2.330E-05 \\
	 \textbf{General PD} &&&\\
	 MDS-UPDRS-1& 3.648E-03 & 2.545E-05 & \textbf{2.418E-02} \\
	 MDS-UPDRS-2 & 6.627E-03 & 3.092E-04 & 2.503E-03 \\
	 MDS-UPDRS-3 & 2.502E-03 & 1.405E-07 & 2.590E-06 \\
	 T-MDS-UPDRS& \textbf{9.561E-03} & 2.809E-04 & 3.607E-03 \\
	 \br
	\end{tabular}
  \end{indented}
\end{table}

Table~\ref{stat2} tabulates the medians and the p-values of the Kruskall-Wallis statistical test to study statistical differences in the growth of variables across layers between individuals belonging to the different variable-communities. Variables with negative directions are denoted by an asterisk (*). To account for Type I errors due to multiple comparisons the Benjamini-Hochberg False Discovery Rate method \cite{BEN95} is used. This gives us an adjusted significance level for each of the p-values $\alpha_{adjusted} = \frac{\alpha \times i}{n_c}$, where $n_c$ is the total number of comparisons, and $i$ is the rank of the p-value. Significance value $\alpha$ is chosen to be $0.05$.  Total number of comparisons $n_c = N_t \times N_v = 3 \times 16 = 46$, where $N_t$ is the number of year-differences, and $N_v$ is the number of variables. Comparisons that do not meet the significance criteria are highlighted in bold in the table. A majority of the values are below their adjusted significance level, suggesting that there exist significant differences in variable progression of the different trajectory-communities.

\subsection{Trajectory Clustering Across an Independent Outcome Variable}
\label{pdx}

Often, medical practitioners may be interested in studying the evolution of disease variables as a function of an outcome variable (a clinical test or variable known to be indicative of specific aspects of disease progression that are of interest). This may, in particular, be useful when complete temporal data about a patient is unavailable, however comprehensive data of evolution of a specific variable is available instead.  Defining layers through values of an outcome variable, allows us to identify subtypes based on trajectories through progression of an outcome variable.

\begin{figure*}[htbp!]
	\begin{center}
		\includegraphics[width=0.95\linewidth]{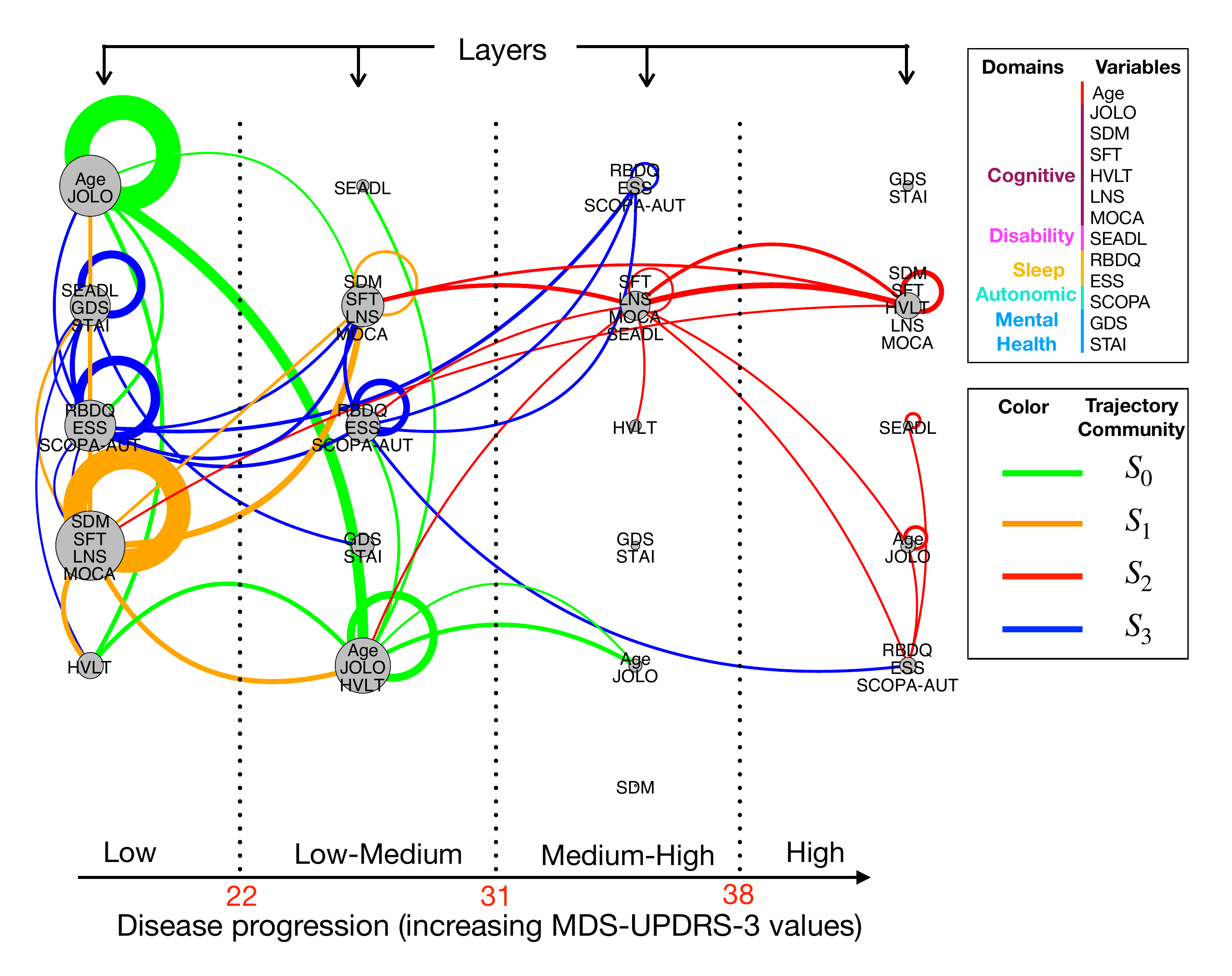}
	\end{center}
	\caption{Trajectory clustering across the outcome variable - baseline MDS-UPDRS3 / PD3 values. The layers represent quartiles of the outcome variable raw values. The boundary values that determine the layers are denoted in red. Each node represents a variable-community consisting of variables and patients. The size of the node denotes the number of patient-years associated with the community. Each patient can contribute upto 5 times (one for each year) to a variable-community. The variables are color-coded by domain in the legend. Patient trajectories directed from the baseline year to the year4 are clustered using the TC algorithm. These trajectories progress in time and hence, are not required to pass through all the outcome variable layers. The trajectory-communities (subtypes) are color-coded. The thickness of a colored edge denotes the number of patients along that edge in the corresponding trajectory-community. Number of people in each trajectory-community are as follows:  green - 47, orange - 44, red - 58, blue - 48. }
	\label{PD_1}
\end{figure*}

\begin{figure*}[htbp!]
	\begin{center}
		\includegraphics[width=0.95\linewidth]{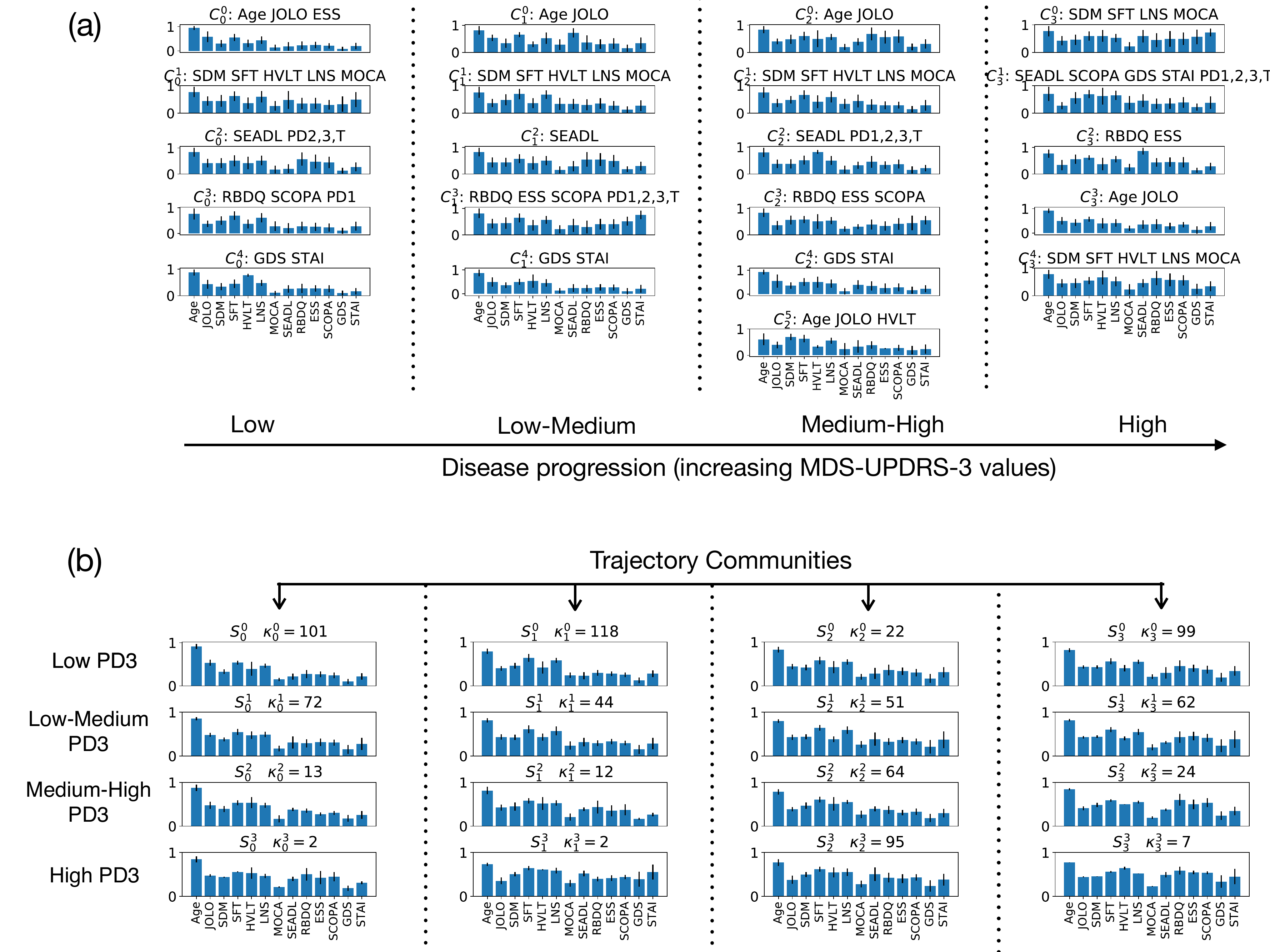}
	\end{center}
	\caption{(a) Variable-community profiles in each layer. Here, the layers represent values of the outcome variable MDS-UPDRS-3. Each panel represents a variable-community comprising of patient-years and variables. The variables in each variable-community are listed above each panel, and are used to identify the variable-communities. The panels show the average $z$ value of all patient-years in that variable-community. (b)Trajectory profiles of the 4 trajectory-communities. The $i^{th}$ vertical set of panels corresponds to a trajectory community $S_i$. Within each column, the four panels arranged from top to bottom show the average profile in layers low, low-mid, mid-high, and high. Each individual panel shows the average $z$ values across all variables of $\kappa$ patient-years in a trajectory-community in a layer. All error bars are given by the standard deviation. }
	\label{PD_2}
\end{figure*}

MDS-UPDRS-3 (divided into quartiles on the x axis) is known to be indicative of PD progression and was chosen as the outcome variable. In figure~\ref{PD_1}, four disease progression trajectory-subtypes across outcome variable layers are identified. The layers are defined by raw baseline MDS-UPDRS-3 values as follows: layer low contains values below $22.0$, layer low-mid contains values between $22.0$ and $31.0$, layer mid-high contains values between $31.0$ and $38.0$, and layer high contains values above $38.0$. Dependent variables are chosen to be the same as in section \ref{timex}, with the exception of General PD severity variables. This method is generalizable to selection of any outcome variable. 

Figure~\ref{PD_1} shows the subtypes identified by the trajectory clustering algorithm. The run time of the TC algorithm on the multi-layer network stacked across outcome variables is $1.7213s$. The size of the nodes in figure~\ref{PD_1} indicate the number of patient-years in that variable-community, i.e., one patient may contribute up to $5$ times to a certain community, one for each timepoint. This implies that a patient may, in consecutive years, remain in the same variable-community, or transition to a different variable-community in the same layer, or transition to a different variable-community in a different layer.

Figure~\ref{PD_2}(a) shows the variable-community profile, i.e., average $z$ values of all the patients in each community in each layer with errors given by their standard deviation. The position of variable-community profiles corresponds to the node in figure~\ref{PD_1}. In figure~\ref{PD_2}(a), one can see that the variables denoting a variable-community have a correspondingly higher relative $z$ value. In figure~\ref{PD_2}(b) shows the average trajectory profile as calculated in \ref{tc_pr} amongst for each trajectory-community across each layer. 

As seen in figure~\ref{PD_1} and figure~\ref{PD_2}(b),The green trajectory-community $S_0$ with $47$ patients largely remains in low to low-medium PD-3 layers, with $101$ patient-years in layer low, and $72$ patient-years in layer low-medium. Individuals in this trajectory-community start with low disease state (as measured by PD-3), and show slow disease progression over the years, staying in the low-medium range of disease progression. 
The orange trajectory-community $S_1$ with $44$ patient-years is the least affected. Patients are most severely affected in the Cognitive domain, however they remain largely in the low disease layer.
The red trajectory-community $S_2$ with $58$ starts in a relatively high disease state, and consistently gets worse, whereas the blue trajectory-community $S_3$ with $48$ starts relatively healthy but with severe disease progression into mid-high and high layers of disease severity measured through PD-3.

\section{Conclusion}

This work introduces a novel algorithm for the identification of subtypes based on relationships between trajectories through a multi-layer network. Multidimensional clinical datasets are often not used to their full potential due to the complexities of a cohesive analysis. This work extracts disease variables that co-express with high severity at different stages of disease progression. Then it extracts trajectories of progression through these variable-communities, i.e., through sets of high severity variables. Lastly, it identifies disease subtypes through clustering patient trajectories. Additionally, it promotes second-order comparisons in the calculation of $\eta$ i.e., correlations across time between two trajectories that do not have overlapping edges but interact with common neighboring trajectories. The agreement of variable-clusters with the domain-structure e.g. Cognitive, Autonomic etc., as well as the statistical analysis, both validate the success of this approach.

Parkinson's Disease has a multitude of clinical variables that interact in complex manners that vary through stages of the disease. A number of studies have identified Parkinson's subtypes based on baseline characteristics \cite{5,20,22}. In contrast, our novel algorithm uses longitudinal data (or patient trajectory over time) to identify disease subtypes through TC. In other words, our method accounts for both disease variable values as well as their progression patterns as a patient progresses through the different layers of the multi-layer stacked-bipartite network. While Parkinson's Disease, being both multivariate and progressive, served to demonstrate the effectiveness of the TC algorithm, the algorithm can, in principle, be easily extended to include non-clinical features, such as genetic or fluid biomarkers, as well as to other datasets.

The TC algorithm identifies subtypes through clustering patient trajectories across layers. This work presents two methods of  analysis and visualization of disease progression: Parkinson's disease trajectory through time, and Parkinson's disease trajectory through an outcome variable (representative of disease progression). Our second-order TC algorithm emphasizes the dynamical aspect of disease progression in addition to the static properties of the associated variables at every stage, contrary to several earlier data-driven methods in medicine that emphasize one or the other. Results are presented in an interpretable visualization that is easily accessible to and comprehensible by medical practitioners in contrast with black box methods in machine learning. 

The TC algorithm is a data-driven network-based method for detection of patient subtype. Like other data-driven methods, this analysis is limited by the availability and quality of the database. The number of variables provided in the database is not exhaustive in the context of Parkinson’s disease. As larger datasets are made available, such results are likely to be more informative and robust. Additionally, in applying a data-driven approach to medical data, important medical decisions must always be made in conjunction with medical expertise. An advantage as well as caveat of this approach is that trajectories are tracked through variable sets, and not through individual variables. Future directions of such work would naturally include extension to other types of medical data (genetic \cite{17}, biomarker etc.), as well as extension to other types of time-evolving and heterogenous datasets. Additional directions of potential interest include studying disease evolution through other outcome variables, as well as treatment-based modifications to the algorithm where effects of treatment are expected during gathering of data. By the nature of such a data-driven method, this approach may not be appropriate for datasets with high variability, low quantity, or data with inconsistent temporal sampling.
 
Parkinson's disease is a highly variable disease with a long onset time. Knowledge of which cluster a new patient belongs to in the baseline year, would allow medical practitioners to predict their subtype and corresponding trajectory, including the type and rate of disease progression. This could open up new and exciting avenues in the field of personalized medicine. Moreover, prediction of disease progression will improve prognostic counseling, a problem commonly encountered by clinicians, by highlighting predicted disease features. It will also support them in seeking more aggressive treatment for patients predicted to display rapid disease progression. Thus, this multi-layer network-based TC approach harnesses data to provide interpretable solutions in the field of early, predictive medicine.

\section*{Data and Code Availability}
Data was obtained from PPMI, a public-private partnership funded by the Michael J. Fox Foundation for Parkinson’s Research and funding partners, including abbvie, Allergan, Avid Radiopharmaceuticals, Biogen, Biolegend, Bristol-Myers Squibb, Celgene, Denali, GE Healthcare, Genentech, gsk, Lilly, Pfizer, Merck, MSD, Lundbeck, Piramal, Prevail Therapeutics, Roche, Sanofi Genzyme, Servier, Takeda, Teva, Ucb, Verily, Voyager Therapeutics and Golub Capital. There are no patents, products in development or marketed products to declare. Python code is available at \textit{www.github.com/chimeraki/Multilayer-Trajectory-Clustering}

\ack{
	The author wishes to sincerely thank Prof. Michelle Girvan, Dr. Lisa M. Shulman, MD and Dr. Rainer von Coelln, MD for helpful discussions and suggestions.}

\section*{References}
\bibliographystyle{unsrt}
\bibliography{res_st}

\appendix
\section{Properties of  second-order Trajectory Clustering}
\label{timex}

The  second-order TC algorithm satisfies three important properties:

\begin{figure}[htbp!]
	\begin{center}
		\includegraphics[width=0.95\linewidth]{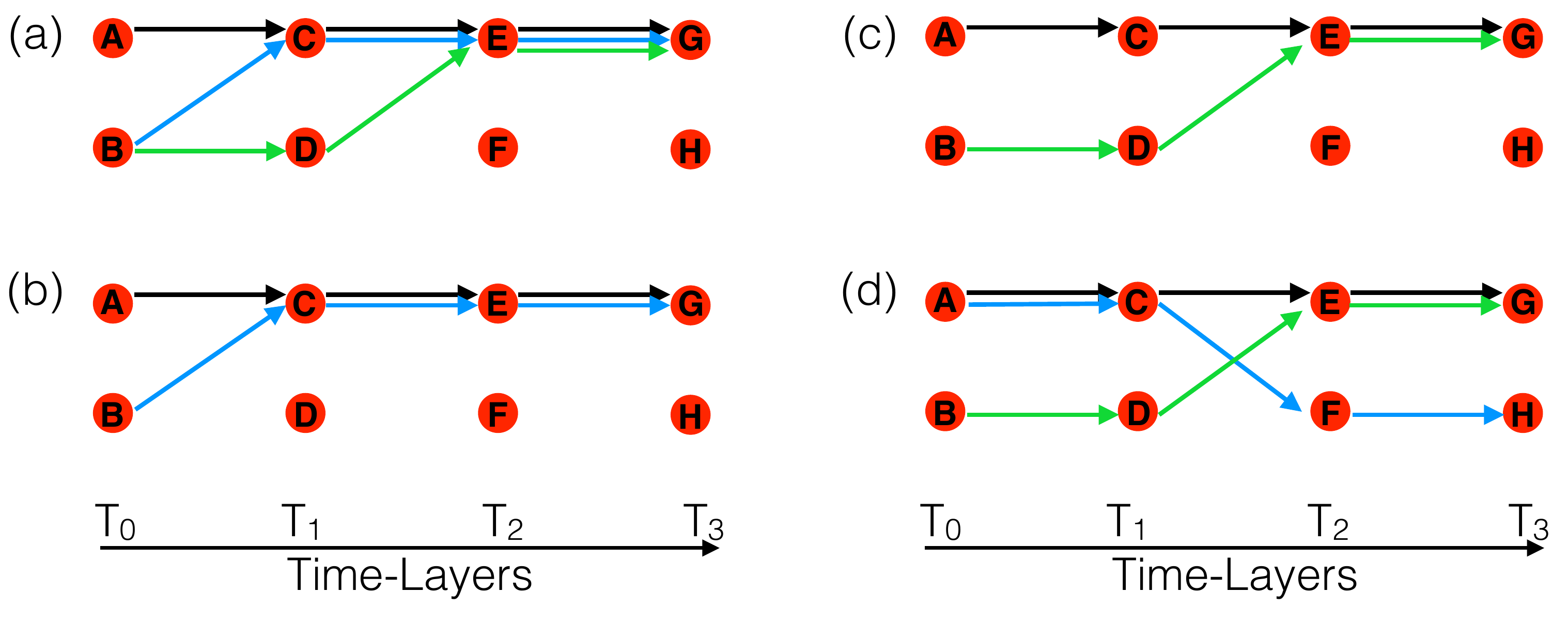}
	\end{center}
	\caption{(a), (b), (c) and (d) are examples of three different trajectories (black, green, blue) going across a temporal network. Each of the nodes represent a variable-community and layers across the x-axis represent time. At each timepoint, there are two variable-communities (A,B), (C,D), (E,F) and (G,H) respectively.}
	\label{networks}
\end{figure}

\begin{enumerate}
	\item Summation: Multiple trajectory pairs with overlapping edges passing through the same node pair reinforce node closeness between them. For instance, consider nodes $A$ and $B$ in figure~\ref{networks}(a). Trajectories $T_{black}$ and $T_{blue}$ (blue) have two overlapping edges $C-E$ and $E-G$. Additionally, $T_{black}$ and $T_{green}$ also have the same overlapping edge $E-G$. Hence, $cl(A,B) \propto 3$ (without normalization), i.e., the contribution of each overlapping trajectory-pair is summed in calculating trajectory similarities. Comparing figure~\ref{networks} (a),(b), and (c) yields $cl(A,B)_a > cl(A,B)_b > cl(A,B)_c$, where the subscript denotes the panel.
	\item Temporal Independence: The algorithm weighs all edges equally. An overlapping edge from layer $T_0$ to layer $T_1$ has the same weight as an overlapping edge from layer $T_2$ to layer $T_3$. 
	\item Temporal Equivalence: Trajectory similarity $\eta$ is calculated based on future as well as past overlap of trajectory pairs. For instance, in figure~\ref{networks}(d), trajectories passing through $C$ and $D$ have exactly one overlapping edge (edge $E-G$ traversed by the black and blue trajectories in the `future', i.e., after the trajectories pass through nodes $C$ and $D$ respectively). Now, nodes $E$ and $F$ also have exactly one overlapping edge (edge $A-C $ traversed by the black and green trajectories in the `past', i.e., before the trajectories pass through nodes $E$ and $F$ respectively). However in figure~\ref{networks}(d), $cl(C,D) = cl(E,F)$, i.e., node closeness is independent of the temporal-occurrence of the overlap. 
\end{enumerate}
These properties ensure that node closeness, and hence trajectory similarity, as described in the main text is truly representative of the second-order similarity between patient profiles, and that similarities in disease progression at all timepoints are equally considered.

\end{document}